\documentclass{article}
\usepackage{amssymb}
\usepackage{amsmath}

\input{tcilatex}
\begin{document}

\begin{center}

%\bigskip

{\Huge Solvable and/or integrable many-body models on a circle}

$\bigskip $

\textbf{Oksana Bihun}$^{\ast 1}$ and \textbf{Francesco Calogero}$%
^{+2}\bigskip $

$^{\ast }$Department of Mathematics, Concordia College at Moorhead, MN, USA

$^{+}$Physics Department, University of Rome \textquotedblleft La Sapienza"
and Istituto Nazionale di Fisica Nucleare, Sezione di Roma

$^{1}$obihun@cord.edu

$^{2}$francesco.calogero@roma1.infn.it, francesco.calogero@uniroma1.it

\bigskip

\textbf{Abstract}
\end{center}

Various many-body models are treated, which describe $N$ points confined to
move on a plane circle. Their Newtonian equations of motion ("accelerations
equal forces") are \textit{integrable}, i. e. they allow the \textit{explicit%
} exhibition of $N$ \textit{constants of motion} in terms of the dependent
variables and their time-derivatives. Some of these models are moreover 
\textit{solvable} by purely algebraic operations, by (explicitly
performable) quadratures and, finally, by functional inversions. The
techniques to manufacture these models are not new; some of these models are
themselves new; others are reinterpretations of known models.

\section{Introduction}

The investigation of the time evolution of an arbitrary number $N$ of
point-particles the dynamics of which is determined by Newtonian equations
of motion ("accelerations equal forces") is of course a fundamental topic in
physics and mathematics. The identification in this context of models 
\textit{amenable to exact treatments} is a major area of research in
mathematical physics and applied mathematics, having a centuries-old history
and having been boosted by developments in the last few decades, which also
impacted several areas of physics beyond mechanics and many fields of pure
mathematics. An interesting related development which is now becoming of
interest is the study of such models in which the motion is restricted to
lie on an \textit{a priori} prescribed manifold: see for instance \cite%
{AM1959} \cite{CG1993} \cite{H2011} \textbf{\cite{ET2012}. }In this paper we
make some initial, simple steps in this direction by focussing on various
many-body models describing the evolution of $N$ points whose positions on a
plane are characterized by $N$ \textit{unit} 2-vectors, thereby forcing
their motion to be confined to \textit{a circle of unit radius centered at
the origin}. All these models are characterized by \textit{Newtonian}
equations of motion: accelerations equal forces, which in these models are
of \textit{one-body}, \textit{two-body} or, in some cases, \textit{many-body}
type, and might depend on the velocities of the moving particles in addition
to their positions. All these models are \textit{autonomous:} their
equations of motion are time-independent. They are \textit{all amenable to
exact treatments}: in particular they \textit{all} allow the explicit
identification of $N$ \textit{constants of motion} in terms of the $N$
dependent variables and their $N$ time-derivatives (for terminological
simplicity we hereafter call such models \textit{integrable}). In some cases
their \textit{initial-value problems} can be moreover \textit{solved} by
(explicitly performable) quadratures and subsequent functional inversions,
preceded by purely algebraic operations, such as solving systems of \textit{%
linear} constant-coefficients ODEs, or (equivalently) evaluating the $N$ 
\textit{eigenvalues} of known (time-dependent) $N\times N$ matrices or
(equivalently) the $N$ \textit{zeros} of known (time-dependent) polynomials
of degree $N$ (for terminological simplicity we hereafter call such models 
\textit{solvable}). The techniques to manufacture these models are \textit{%
not new}; some of these models are themselves \textit{new}; others are
essentially \textit{reinterpretations of known models}. The dynamics of
these models are not analyzed in detail; but in some cases the main features
of their behavior are ascertained, for instance for \textit{isochronous}
models the time evolution of which is \textit{isochronous} (i. e., \textit{%
completely periodic} with a \textit{fixed} period independent of the initial
data), or for models \textit{all} motions of which are \textit{multiply
periodic}.

The equations of motion of the $N$-body problems treated below are listed
with minimal comments in the following Section 2, to facilitate the hasty
reader wishing to get an immediate idea of the findings reported in this
paper. These results are then proven in the subsequent Section 3: the titles
of its subsections indicate case-by-case the techniques employed to arrive
at the relevant results. Finally, a terse Section 4 entitled "Outlook"
outlines possible developments, to be eventually reported in other papers.
Some mathematical details are confined to two Appendices.

\section{Many-body models on a circle amenable to exact treatments}

In the following subsections we display, with minimal comments, various $N$%
-body problems of Newtonian type ("accelerations equal forces") describing
motions on a circle and amenable to exact treatments (detailed in the
following Section 3). But we provide firstly a terse subsection devoted to
notation.

\subsection{Notations}

The models under consideration generally feature $N$ points moving in a
plane. We identify these $N$ points by 3-vectors $\vec{r}_{n}$, $n=1,2,...,N$
for which we use the following 3-dimensional notation:%
\begin{equation}
\vec{r}_{n}\equiv \left( \cos \theta _{n},~\sin \theta _{n},~0\right) \equiv
\left( x_{n},~y_{n},~0\right) ~.  \label{rn}
\end{equation}

Hereafter $N$ is an \textit{arbitrary positive integer} (generally $N\geq 2$%
) and indices such as $n,$ $m,$ $\ell $ run over the \textit{positive
integers} from $1$ to $N$ (unless otherwise explicitly indicated).

Clearly these vectors $\vec{r}_{n}$ have \textit{unit} length, 
\begin{subequations}
\begin{equation}
\vec{r}_{n}\cdot \vec{r}_{n}=1~.  \label{rnunit}
\end{equation}%
Throughout this paper the dot sandwiched among two vectors denotes the
standard \textit{scalar} product, so that for instance%
\begin{equation}
\vec{r}_{n}\cdot \vec{r}_{m}=\cos \left( \theta _{n}-\theta _{m}\right) ~.
\label{rndotrm}
\end{equation}%
It is moreover convenient to introduce the \textit{unit} vector $\hat{z}$
orthogonal to the $xy$-plane, 
\end{subequations}
\begin{equation}
\hat{z}\equiv \left( 0,~0,~1\right) ~,  \label{zhat}
\end{equation}%
and to denote by the "wedge" symbol $\wedge $ the standard (3-dimensional) 
\textit{vector} product, so that 
\begin{subequations}
\begin{equation}
\hat{z}\wedge \vec{r}_{n}=-\vec{r}_{n}\wedge \hat{z}=\left( -\sin \theta
_{n},~\cos \theta _{n},~0\right) ~,
\end{equation}%
\begin{equation}
\left( \hat{z}\wedge \vec{r}_{m}\right) \cdot \vec{r}_{n}=\left( \vec{r}%
_{m}\wedge \vec{r}_{n}\right) \cdot \hat{z}=\sin \left( \theta _{n}-\theta
_{m}\right) ~.  \label{rmwedgerndotzhat}
\end{equation}

Hereafter we deal with time-dependent vectors 
\end{subequations}
\begin{equation}
\vec{r}_{n}\left( t\right) \equiv \left( \cos \theta _{n}\left( t\right)
,~\sin \theta _{n}\left( t\right) ,~0\right) ~,
\end{equation}%
and superimposed dots indicate derivatives with respect to the time variable 
$t$ so that, for instance, 
\begin{subequations}
\begin{equation}
\overset{\cdot }{\vec{r}}_{n}=\dot{\theta}_{n}~\left( -\sin \theta
_{n},~\cos \theta _{n},~0\right) =\dot{\theta}_{n}~\hat{z}\wedge \vec{r}%
_{n}~,  \label{rndot}
\end{equation}%
\begin{eqnarray}
\overset{\cdot \cdot }{\vec{r}}_{n} &=&\ddot{\theta}_{n}~\left( -\sin \theta
_{n},~\cos \theta _{n},~0\right) -\dot{\theta}_{n}^{2}~\left( \cos \theta
_{n},~\sin \theta _{n},~0\right)  \notag \\
&=&\ddot{\theta}_{n}~\hat{z}\wedge \vec{r}_{n}-\dot{\theta}_{n}^{2}~\vec{r}%
_{n}~.  \label{rndotdot}
\end{eqnarray}%
Note that here we omitted, for notational simplicity, to indicate \textit{%
explicitly} the time-dependence of the quantities appearing in these $N$
equations; we will often do this below without repeating this warning.

Several other identities are reported in Appendix A: they are useful to
obtain the results reported below, but are not necessary to understand the
findings reported in the following subsections.

\subsection{Two models obtained via techniques of generalized Lagrangian
interpolation}

\textit{First model}: 
\end{subequations}
\begin{subequations}
\label{ManyBodyForcesModel}
\begin{eqnarray}
&&\mu _{n}~\overset{\cdot \cdot }{\vec{r}}_{n}=-\mu _{n}~\left( \overset{%
\cdot }{\vec{r}}_{n}\cdot \overset{\cdot }{\vec{r}}_{n}\right) ~\vec{r}_{n} 
\notag \\
&&+\hat{z}\wedge \vec{r}_{n}~\left\{ \left[ \mu _{n}~\left( \overset{\cdot }{%
\vec{r}}_{n}\cdot \overset{\cdot }{\vec{r}}_{n}\right) +\eta _{n}~\left( 
\vec{r}_{n}\wedge \overset{\cdot }{\vec{r}}_{n}\right) \cdot \hat{z}\right]
~\sum_{\ell =1,~\ell \neq n}^{N}\left[ \frac{\left( \vec{r}_{\ell }\cdot 
\vec{r}_{n}\right) }{\left( \vec{r}_{\ell }\wedge \vec{r}_{n}\right) \cdot 
\hat{z}}\right] \right.  \notag \\
&&\left. +\left[ \left( \vec{r}_{n}\wedge \overset{\cdot }{\vec{r}}%
_{n}\right) \cdot \hat{z}\right] \sum_{\ell =1,~\ell \neq n}^{N}\left[ \frac{%
\sigma _{n}\left( \underline{\vec{r}}\right) }{\sigma _{\ell }\left( 
\underline{\vec{r}}\right) }~\frac{\mu _{\ell }~\left( \vec{r}_{\ell }\wedge 
\overset{\cdot }{\vec{r}}_{\ell }\right) \cdot \hat{z}+\eta _{\ell }}{\left( 
\vec{r}_{\ell }\wedge \vec{r}_{n}\right) \cdot \hat{z}}\right] \right\} ~,
\label{ManyBodyForcesOnCircle}
\end{eqnarray}%
\begin{equation}
\sigma _{n}\left( \underline{\vec{r}}\right) =\dprod\limits_{\ell =1,~\ell
\neq n}^{N}\left[ \left( \vec{r}_{\ell }\wedge \vec{r}_{n}\right) \cdot \hat{%
z}\right] ~.  \label{sigman}
\end{equation}

\textit{Second model}: 
\end{subequations}
\begin{eqnarray}
&&\mu _{n}~\overset{\cdot \cdot }{\vec{r}}_{n}==-\mu _{n}~\left( \overset{%
\cdot }{\vec{r}}_{n}\cdot \overset{\cdot }{\vec{r}}_{n}\right) ~\vec{r}_{n} 
\notag \\
&&+\sum_{\ell =1,~\ell \neq n}^{N}\left\{ \left[ \left( \vec{r}_{\ell
}\wedge \vec{r}_{n}\right) \cdot \hat{z}\right] ^{-1}~\left\{ \left[ \left( 
\vec{r}_{n}\wedge \overset{\cdot }{\vec{r}}_{n}\right) \cdot \hat{z}\right] ~%
\left[ \mu _{\ell }~\left( \vec{r}_{\ell }\wedge \overset{\cdot }{\vec{r}}%
_{\ell }\right) \cdot \hat{z}+\eta _{\ell }\right] \right. \right.  \notag \\
&&\left. \left. +\left[ \mu _{n}~\left( \vec{r}_{n}\wedge \overset{\cdot }{%
\vec{r}}_{n}\right) \cdot \hat{z}+\eta _{n}\right] ~\left[ \left( \vec{r}%
_{\ell }\wedge \overset{\cdot }{\vec{r}}_{\ell }\right) \cdot \hat{z}\right]
\right\} ~\left( \vec{r}_{\ell }\wedge \vec{r}_{n}\right) \right\} ~.
\label{TwoBodyForcesOnCircle}
\end{eqnarray}%
In these Newtonian equations $\mu _{n}$ and $\eta _{n}$ are $2N$ arbitrary
constants, and for the rest of the notation see Subsection 2.1; note in
particular the property (\ref{rnunit}), implying that the $N$ vectors $\vec{r%
}_{n}$ have \textit{unit} modulus, hence that the $N$ points whose time
evolution is determined by these equations of motion are constrained to move
on the circle of \textit{unit} radius centered at the origin of the
Cartesian plane.

These equations of motion are \textit{covariant}, implying that the
corresponding $N$-body problems are \textit{rotation-invariant}.

These two $N$-body problems are both \textit{integrable}: they possess $N$ 
\textit{constants of motion}, the explicit expressions of which in terms of
the vectors $\vec{r}_{n}$ and their time-derivatives $\overset{\cdot }{\vec{r%
}}_{n}$ are displayed in the following Subsection 3.1. The equations of
motion of the first, (\ref{ManyBodyForcesOnCircle}), of these two models
feature \textit{many-body} forces due to the presence in their right-hand
("forces") sides of the quantities $\sigma _{n}\left( \underline{\vec{r}}%
\right) $, see (\ref{sigman}), but \textit{their initial-value problem is} 
\textit{solvable by purely algebraic operations}; nevertheless their time
evolution can be quite complicated (detailed analyses are not performed in
this paper; the fact that \textit{solvable} models can exhibit quite
complicated dynamics is of course well known, see for instance the papers
where a 3-body model is studied the time evolution of which is highly
nontrivial in spite of the fact that its Aristotelian equations of
motion---"velocity equal forces"---are quite neat and that its initial-value
problem can be reduced to solving a single algebraic equation \cite{CGSS}).%
\textit{\ }

\subsection{Two \textit{solvable} models obtained via a reinterpretation of
known models}

The \textit{first model} is merely a transcription of the \textit{solvable}
"Sutherland model", see Subsection 3.2. It reads as follows:%
\begin{equation}
\overset{\cdot \cdot }{\vec{r}}_{n}=-\left( \overset{\cdot }{\vec{r}}%
_{n}\cdot \overset{\cdot }{\vec{r}}_{n}\right) ~\vec{r}_{n}+g^{2}~\hat{z}%
\wedge \vec{r}_{n}~\sum_{\ell =1,~\ell \neq n}^{N}\left\{ \frac{\vec{r}%
_{n}\cdot \vec{r}_{\ell }}{\left[ \left( \vec{r}_{\ell }\wedge \vec{r}%
_{n}\right) \cdot \hat{z}\right] ^{3}}\right\} ~.  \label{2a}
\end{equation}%
Here $g$ is an \textit{arbitrary} "coupling constant", and the rest of the
notation is, we trust, clear (see Subsection 2.1).

The \textit{second model} is also merely a transcription of a well-known 
\textit{solvable} model ("of goldfish type"), see Subsection 3.2. It reads
as follows:%
\begin{eqnarray}
&&\overset{\cdot \cdot }{\vec{r}}_{n}=-\left( \overset{\cdot }{\vec{r}}%
_{n}\cdot \overset{\cdot }{\vec{r}}_{n}\right) ~\vec{r}_{n}+g_{0}~\hat{z}%
\wedge \vec{r}_{n}+g_{1}~\overset{\cdot }{\vec{r}}_{n}  \notag \\
&&+\hat{z}\wedge \vec{r}_{n}~\sum_{\ell =1,~\ell \neq n}^{N}\left\{ \frac{2~%
\overset{\cdot }{\vec{r}}_{n}\cdot \overset{\cdot }{\vec{r}}_{\ell }+g_{2}~%
\left[ \left( \overset{\cdot }{\vec{r}}_{n}\wedge \vec{r}_{\ell }+\overset{%
\cdot }{\vec{r}}_{\ell }\wedge \vec{r}_{n}\right) \cdot \hat{z}\right]
+g_{3}~\vec{r}_{n}\cdot \vec{r}_{\ell }}{\left( \vec{r}_{\ell }\wedge \vec{r}%
_{n}\right) \cdot \hat{z}}\right\} ~.  \notag \\
&&  \label{2b}
\end{eqnarray}%
Here the $4$ constants $g_{0},$ $g_{1},$ $g_{2}$ and $g_{3}$ are \textit{%
arbitrary} constants, and the rest of the notation is, we trust, clear (see
Subsection 2.1).

These equations of motion are \textit{covariant}, implying that the
corresponding $N$-body problems are \textit{rotation-invariant}.

\subsection{Two $N$-body problems on a circle obtained by changes of
dependent variables}

These two \textit{solvable} models are merely transcriptions of two
well-known one-dimensional \textit{solvable} models, see Subsection 3.3. The 
\textit{first model} reads as follows: 
\begin{subequations}
\begin{eqnarray}
&&\overset{\cdot \cdot }{\vec{r}}_{n}=-\left( \overset{\cdot }{\vec{r}}%
_{n}\cdot \overset{\cdot }{\vec{r}}_{n}\right) ~\vec{r}_{n}-\hat{z}\wedge 
\vec{r}_{n}~\left\{ 2~\left[ \left( \overset{\cdot }{\vec{r}}_{n}\cdot 
\overset{\cdot }{\vec{r}}_{n}\right) ~\frac{y_{n}}{x_{n}}\right] \right. 
\notag \\
&&\left. +4~x_{n}~y_{n}-x_{n}^{5}~\sum_{\ell =1,~\ell \neq n}^{N}\left[ 
\frac{y_{\ell }}{\left( \vec{r}_{\ell }\wedge \vec{r}_{n}\right) \cdot \hat{z%
}}\right] ^{3}\right\} ~.  \label{CalCircle}
\end{eqnarray}%
Here $x_{n}\equiv \cos \theta _{n}$ and $y_{n}\equiv \sin \theta _{n}$ are
the two Cartesian components in the plane of the vector $\vec{r}_{n},$ see (%
\ref{rn}).

This model is \textit{isochronous} with period $\pi $,%
\begin{equation}
\vec{r}_{n}\left( t\pm \pi \right) =\vec{r}_{n}\left( t\right) ~.
\end{equation}

The \textit{second model} reads as follows: 
\end{subequations}
\begin{eqnarray}
&&\overset{\cdot \cdot }{\vec{r}}_{n}=-\left( \overset{\cdot }{\vec{r}}%
_{n}\cdot \overset{\cdot }{\vec{r}}_{n}\right) ~\vec{r}_{n}-\hat{z}\wedge 
\vec{r}_{n}~\left\{ 2~\left[ \left( \overset{\cdot }{\vec{r}}_{n}\cdot 
\overset{\cdot }{\vec{r}}_{n}\right) ~\frac{y_{n}}{x_{n}}\right] \right. 
\notag \\
&&\left. +x_{n}~y_{n}-x_{n}~\sum_{\ell =1,~\ell \neq n}^{N}\left\{ \frac{%
2~+x_{n}^{2}~x_{\ell }^{2}}{x_{\ell }~\left[ \left( \vec{r}_{\ell }\wedge 
\vec{r}_{n}\right) \cdot \hat{z}\right] }\right\} \right\} ~.
\label{GoldCircle}
\end{eqnarray}%
Here $x_{n}\equiv \cos \theta _{n}$ and $y_{n}\equiv \sin \theta _{n}$ are
again the two Cartesian components in the plane of the vector $\vec{r}_{n},$
see (\ref{rn}).

\textit{All} solutions of this model are \textit{multiply periodic}, see
Subsection 3.3.

Note that---in contrast to the equations of motions reported in the two
preceding subsections---those displayed herein, (\ref{CalCircle}) and (\ref%
{GoldCircle}), are\textit{\ not} written in \textit{covariant} fashion, i.
e. without any explicit appearance of the Cartesian components $x_{n}\equiv
\cos \theta _{n}$ and $y_{n}\equiv \sin \theta _{n}$ of the vector $\vec{r}%
_{n}$; indeed these equations of motion are \textit{not} rotation-invariant,
or equivalently, they are \textit{not} invariant for translations along the
circle (on which the motions take place due to the constraint (\ref{rnunit}%
)).

\section{Proofs}

In the following subsections we substantiate the findings reported in the
preceding Section 2.

\subsection{Solvable and integrable models on the circle manufactured via
techniques of generalized Lagrangian interpolation}

In this subsection we employ the technique to manufacture many-body models
amenable to exact treatments introduced in \cite{C2001} (see in particular
Chapter 3 of this book, entitled "$N$-body problems treatable via techniques
of exact Lagrangian interpolation in spaces of one or more dimensions"). We
begin with a terse review of this method, in the specific case of
one-dimensional space with an appropriate choice of the set of "seeds"
(namely, of the $N$ functions providing the point of departure for the
generalized Lagrangian interpolation approach).

The set of seeds we conveniently take as basis for our treatment are the $N$
functions 
\begin{eqnarray}
&&\left\{ s_{n}\left( \theta \right) \right\} _{n=1}^{N}=\left\{ \exp \left[
i~\left( 2~n-N-1\right) ~\theta \right] \right\} _{n=1}^{N}  \notag \\
&=&\{\exp \left[ i~\left( 1-N\right) ~\theta \right] ,~\exp \left[ i~\left(
3-N\right) ~\theta \right] ,~...  \notag \\
&&...\exp \left[ i~\left( N-3\right) ~\theta \right] ,~\exp \left[ i~\left(
N-1\right) ~\theta \right] \}~.  \label{seeds}
\end{eqnarray}

\textit{Remark 3.1.1}. These exponential functions with \textit{imaginary}
argument are \textit{complex}, but clearly this set of seeds could be
replaced without significant changes by an equivalent set featuring instead
sines and cosines of \textit{real} arguments. The use of exponentials merely
facilitates some of the following developments. Likewise the factor $2$ in
the argument of these functions has been introduced merely to yield neater
versions of the equations of motions that will be obtained, see below. The
fact that these seeds are invariant under the transformation $\theta
\Rightarrow \theta +2\pi $ suggests to interpret the variable $\theta $ as
an \textit{angle} in the plane. $\blacksquare $

We then consider a function $f\left( \theta \right) $ representable as a 
\textit{linear} superposition of these $N$ seeds, 
\begin{subequations}
\label{GenLag}
\begin{equation}
f\left( \theta \right) =\sum_{n=1}^{N}\left[ h_{n}~s_{n}\left( \theta
\right) \right] ~,  \label{fhn}
\end{equation}%
where the $N$ coefficients $h_{n}$ are \textit{a priori} arbitrary numbers.
And we denote with $f_{n}$ the $N$ values that this function takes at the $N$
(\textit{arbitrarily assigned}) "nodes" $\theta =\theta _{n}$,%
\begin{equation}
f_{n}=f\left( \theta _{n}\right) ~;  \label{fn}
\end{equation}%
and we display the representation of this function in terms of these $N$
values, via the ("generalized Lagrangian interpolation") formula%
\begin{equation}
f\left( \theta \right) =\sum_{n=1}^{N}\left[ f_{n}~q^{\left( n\right)
}\left( \theta ~\left\vert \underline{\theta }\right. \right) \right] ~.
\label{fqn}
\end{equation}%
The $N$ "interpolational functions" $q^{\left( n\right) }\left( \theta
~\left\vert \underline{\theta }\right. \right) $ depend on the variable $%
\theta $ and on the $N$ nodes $\theta _{n}$ (hence on the $N$-vector having
these nodes as its components, hereafter denoted as $\underline{\theta }%
\equiv \left( \theta _{1},~\theta _{2},~...,~\theta _{N}\right) $); they are
themselves \textit{linear} superpositions of the seeds $s_{n}\left( \theta
\right) $, to insure consistency among (\ref{fqn}) and (\ref{fhn}); and they
feature the property 
\end{subequations}
\begin{equation}
q^{\left( n\right) }\left( \theta _{m}~\left\vert \underline{\theta }\right.
\right) =\delta _{nm}  \label{qdeltanm}
\end{equation}%
to insure consistency among (\ref{fqn}) and (\ref{fn}) (here and hereafter $%
\delta _{nm}$ is the Kronecker symbol: $\delta _{nm}=1$ if $n=m$, $\delta
_{nm}=0$ if $n\neq m$).

The explicit representation of these interpolational functions $q^{\left(
n\right) }\left( \theta ~\left\vert \underline{\theta }\right. \right) $ in
terms of the $N$ seeds $s_{n}\left( \theta \right) $ and the $N$ nodes $%
\theta _{n}$ reads \cite{C2001} 
\begin{subequations}
\label{Reprqn}
\begin{equation}
q^{(n)}(\theta ~\left\vert \underline{\theta }\right. )=\frac{\Delta (\theta
_{1},\ldots ,\theta _{n-1},\theta ,\theta _{n+1},\ldots ,\theta _{N})}{%
\Delta (\theta _{1},\ldots ,\theta _{N})}~,
\end{equation}%
where 
\begin{equation}
\Delta (\underline{\theta })=\left\vert 
\begin{array}{cccc}
s_{1}(\theta _{1}) & s_{2}(\theta _{1}) & \ldots & s_{N}(\theta _{1}) \\ 
s_{1}(\theta _{2}) & s_{2}(\theta _{2}) & \ldots & s_{N}(\theta _{2}) \\ 
\vdots & \vdots & \ddots & \vdots \\ 
s_{1}(\theta _{N}) & s_{2}(\theta _{N}) & \ldots & s_{N}(\theta _{N})%
\end{array}%
\right\vert ~.
\end{equation}%
This determinant---with the set of seeds (\ref{seeds})---is of Vandermonde
type hence it can be explicitly evaluated, yielding for the interpolational
functions the expression 
\end{subequations}
\begin{equation}
q^{\left( n\right) }\left( \theta ~\left\vert \underline{\theta }\right.
\right) =s_{1}\left( \theta -\theta _{n}\right) ~\dprod\nolimits_{\ell
=1,~\ell \neq n}^{N}\left[ \frac{\exp \left( 2~i~\theta \right) -\exp \left(
2~i~\theta _{\ell }\right) }{\exp \left( 2~i~\theta _{n}\right) -\exp \left(
2~i~\theta _{\ell }\right) }\right] ~.  \label{qn}
\end{equation}

The next step is to introduce the time variable $t$. As in \cite{C2001}, we
assume hereafter that the $N$ seeds $s_{n}\left( \theta \right) $ are
time-independent; we moreover assume the function $f\left( \theta \right) $
to be also time-independent (thereby simplifying the more general treatment
of \cite{C2001}). A time-dependence is only introduced for the nodes $\theta
_{n}\equiv \theta _{n}\left( t\right) ;$ indeed they shall be the dependent
variables of the dynamical systems we manufacture. Of course the fact that
the nodes $\theta _{n}\left( t\right) $ evolve over time entails that the
values $f_{n}$ taken by the function $f\left( \theta \right) $ at these
nodes (see (\ref{fn})) also evolve over time: 
\begin{equation}
f_{n}\equiv f_{n}\left( t\right) =f\left[ \theta _{n}\left( t\right) \right]
~.  \label{fnt}
\end{equation}

We then posit a convenient relation among the time evolution of the $N$
nodes $\theta _{n}\left( t\right) $ and the time evolution of the $N$
quantities $f_{n}\left( t\right) $, by setting%
\begin{equation}
f_{n}\left( t\right) =\rho _{n}\left[ \underline{\theta }\left( t\right) %
\right] ~\dot{\theta}_{n}\left( t\right) +\gamma _{n}\left[ \underline{%
\theta }\left( t\right) \right] ~.  \label{qndot}
\end{equation}%
Here we introduced the $2N$ functions $\rho _{n}\left( \underline{\theta }%
\right) $ and $\gamma _{n}\left( \underline{\theta }\right) $ of the $N$
nodes $\theta _{n}$, that will be assigned later at our convenience (but
note that we forsake---again, for simplicity---the possibility to assign an 
\textit{explicit} time-dependence to these functions, in addition to their
dependence on the $N$ nodes).

The next step is to ascertain the time dependence of the $N$ nodes $\theta
_{n}\equiv \theta _{n}\left( t\right) $ implied by these assignments. To
this end we time-differentiate the relation (\ref{qndot}), getting the
following expressions for the second time-derivatives of the $N$ nodes $%
\theta _{n}\equiv \theta _{n}\left( t\right) $:%
\begin{equation}
\rho _{n}\left( \underline{\theta }\right) ~\ddot{\theta}_{n}=\dot{f}%
_{n}-\sum_{m=1}^{N}\left\{ \left[ \frac{\partial ~\gamma _{n}\left( 
\underline{\theta }\right) }{\partial ~\theta _{m}}+\frac{\partial ~\rho
_{n}\left( \underline{\theta }\right) }{\partial ~\theta _{m}}~\dot{\theta}%
_{n}\right] ~\dot{\theta}_{m}\right\} ~.  \label{thetandotdot}
\end{equation}

Our next step is to evaluate the quantity $\dot{f}_{n},$ which (see (\ref%
{fnt})) reads%
\begin{equation}
\dot{f}_{n}=\frac{\partial ~f\left( \theta _{n}\right) }{\partial ~\theta
_{n}}~\dot{\theta}_{n}~.
\end{equation}%
To evaluate this quantity we can use the finite-dimensional representation
of the differential operator, yielding (for functions which are \textit{%
linear} superpositions of the seeds $s_{n}\left( \theta \right) $, see (\ref%
{GenLag})), the \textit{exact} formula \cite{C2001} 
\begin{subequations}
\begin{equation}
\frac{\partial ~f\left( \theta _{n}\right) }{\partial ~\theta _{n}}%
=\sum_{m=1}^{N}\left[ D_{nm}\left( \underline{\theta }\right) ~f_{m}\right]
~,
\end{equation}%
with the $N\times N$ matrix $D$ defined componentwise as follows \cite{C2001}%
:%
\begin{equation}
D_{nm}\left( \underline{\theta }\right) =\frac{\partial ~q^{\left( m\right)
}(\theta ~\left\vert \underline{\theta }\right. )}{\partial ~\theta }~~~%
\text{evaluated at~~~}\theta =\theta _{n}~,\ 
\end{equation}%
hence in our case (see (\ref{seeds}) and (\ref{Reprqn})) reading 
\end{subequations}
\begin{subequations}
\label{MatrixD}
\begin{equation}
D_{nm}\left( \underline{\theta }\right) =\delta _{nm}~\sum_{\ell =1,~\ell
\neq n}^{N}\cot \left( \theta _{n}-\theta _{\ell }\right) +\left( 1-\delta
_{nm}\right) ~\frac{\sigma _{n}\left( \underline{\theta }\right) }{\sigma
_{m}\left( \underline{\theta }\right) }~\frac{1}{\sin \left( \theta
_{n}-\theta _{m}\right) }~,  \label{Dnm}
\end{equation}%
\begin{equation}
\sigma _{n}\left( \underline{\theta }\right) =\dprod\limits_{\ell =1,~\ell
\neq n}^{N}\left[ \sin \left( \theta _{n}-\theta _{\ell }\right) \right] ~.
\label{sigma}
\end{equation}%
Note that this definition coincides, via (\ref{rmwedgerndotzhat}), with (\ref%
{sigman}).

We therefore conclude that the system (\ref{thetandotdot}) yields the
following set of $N$ Newtonian equations of motion for the dependent
variables $\theta _{n}\equiv \theta _{n}\left( t\right) $: 
\end{subequations}
\begin{eqnarray}
&&\rho _{n}\left( \underline{\theta }\right) ~\ddot{\theta}_{n}=\dot{\theta}%
_{n}~\left[ \rho _{n}\left( \underline{\theta }\right) ~\dot{\theta}%
_{n}+\gamma _{n}\left( \underline{\theta }\right) \right] ~\sum_{\ell
=1,~\ell \neq n}^{N}\left[ \cot \left( \theta _{n}-\theta _{\ell }\right) %
\right]  \notag \\
&&+\dot{\theta}_{n}~\sum_{\ell =1,~\ell \neq n}^{N}\left\{ \frac{\sigma
_{n}\left( \underline{\theta }\right) }{\sigma _{\ell }\left( \underline{%
\theta }\right) }~\frac{\left[ \rho _{\ell }\left( \underline{\theta }%
\right) ~\dot{\theta}_{\ell }+\gamma _{\ell }\left( \underline{\theta }%
\right) \right] }{\sin \left( \theta _{n}-\theta _{\ell }\right) }\right\} 
\notag \\
&&-\sum_{m=1}^{N}\left\{ \left[ \frac{\partial ~\rho _{n}\left( \underline{%
\theta }\right) }{\partial ~\theta _{m}}~\dot{\theta}_{n}+\frac{\partial
~\gamma _{n}\left( \underline{\theta }\right) }{\partial ~\theta _{m}}\right]
~\dot{\theta}_{m}\right\} ~.  \label{GenNbody}
\end{eqnarray}%
Of course to obtain this system of $N$ second-order ODEs we also used (\ref%
{qndot}).

Let us now emphasize that, as a consequence of the way these $N$-body
problems have been manufactured, they are \textit{integrable}$.$ It is
indeed plain that the time independence of the function $f\left( \theta
\right) $ entails (via (\ref{fhn}), (\ref{fn}) and (\ref{qndot})) the
relations 
\begin{subequations}
\label{FirstOrderODEs}
\begin{equation}
\sum_{m=1}^{N}\left\{ h_{m}~s_{m}\left[ \theta _{n}\left( t\right) \right]
\right\} =\rho _{n}\left[ \underline{\theta }\left( t\right) \right] ~\dot{%
\theta}_{n}\left( t\right) +\gamma _{n}\left[ \underline{\theta }\left(
t\right) \right] ~.  \label{hthetadot}
\end{equation}%
Here we have displayed the \textit{time-dependence} of the various
quantities, in order to emphasize the \textit{time-independence} of the $N$
coefficients $h_{m}$, which can actually be evaluated by solving this system
of $N$ \textit{linear }equations, thereby obtaining (via (\ref{Reprqn})) the
following formulas:%
\begin{equation}
h_{m}=q^{\left( m\right) }\left( \vartheta _{m}~\left\vert \underline{\theta 
}\right. \right) ~,~~~~\vartheta _{m}\equiv \frac{i~\log \left[ \rho
_{m}\left( \underline{\theta }\right) ~\dot{\theta}_{m}+\gamma _{m}\left( 
\underline{\theta }\right) \right] }{2~m-N-1}~,  \label{hn}
\end{equation}%
where of course the $N$ nodes $\theta _{m}\equiv \theta _{m}\left( t\right) $
and their $N$ time derivatives $\dot{\theta}_{m}\equiv \dot{\theta}%
_{m}\left( t\right) $ can be evaluated at any arbitrary time $t$. It is thus
plain that the $N$-body systems (\ref{GenNbody}) are \textit{integrable} for
any \textit{arbitrary} assignment of the $2N$ functions $\rho _{m}\left( 
\underline{\theta }\right) $ and $\gamma _{m}\left( \underline{\theta }%
\right) $ of the $N$ dependent variables $\theta _{n}$, with these $N$
quantities $h_{m}$ providing $N$ \textit{constants of motion} given by
explicit (generally nontrivial) expressions in terms of the $N$ nodes $%
\theta _{n}$ and their $N$ time-derivatives $\dot{\theta}_{n}$.

We are still free to assign the $2N$ functions $\rho _{n}\left( \underline{%
\theta }\right) $ and $\gamma _{n}\left( \underline{\theta }\right) .$ There
are two natural choices.

The first one reads simply 
\end{subequations}
\begin{equation}
\rho _{n}\left( \underline{\theta }\right) =\mu _{n}\text{ ,~~~}\gamma
_{n}\left( \underline{\theta }\right) =\eta _{n}~,  \label{rhogammaconstants}
\end{equation}%
with $\mu _{n}$ and $\eta _{n}$ arbitrary \textit{constant} parameters. It
clearly yields (see (\ref{GenNbody})) an $N$-body system characterized by
the following set of Newtonian equations of motion:%
\begin{eqnarray}
&&\mu _{n}~\ddot{\theta}_{n}=\dot{\theta}_{n}~\left( \mu _{n}~\dot{\theta}%
_{n}+\eta _{n}\right) ~\sum_{\ell =1,~\ell \neq n}^{N}\left[ \cot \left(
\theta _{n}-\theta _{\ell }\right) \right]  \notag \\
&&+\dot{\theta}_{n}~\sum_{\ell =1,~\ell \neq n}^{N}\left[ \frac{\sigma
_{n}\left( \underline{\theta }\right) }{\sigma _{\ell }\left( \underline{%
\theta }\right) }~\frac{\left( \mu _{\ell }~\dot{\theta}_{\ell }+\eta _{\ell
}\right) }{\sin \left( \theta _{n}-\theta _{\ell }\right) }~\right] ~.
\label{WithManyBodyForces}
\end{eqnarray}%
Here the functions $\sigma _{n}\left( \underline{\theta }\right) $ of the $N$
nodes $\theta _{m}$ are of course defined by (\ref{sigma}).

The second assignment of the $2N$ functions $\rho _{n}\left( \underline{%
\theta }\right) $ and $\gamma _{n}\left( \underline{\theta }\right) $ is
suggested by the structure of the system (\ref{GenNbody}). It reads%
\begin{equation}
\rho _{n}\left( \underline{\theta }\right) =\mu _{n}\text{~}\sigma
_{n}\left( \underline{\theta }\right) \text{,~\ ~}\gamma _{n}\left( 
\underline{\theta }\right) =\eta _{n}~\sigma _{n}\left( \underline{\theta }%
\right) ~,  \label{rhogmmasigma}
\end{equation}%
where again $\mu _{n}$ and $\eta _{n}$ are arbitrary \textit{constant}
parameters and the functions $\sigma _{n}\left( \underline{\theta }\right) $
are defined as above, see (\ref{sigma}), implying (by logarithmic
differentiation) 
\begin{subequations}
\begin{equation}
\frac{\partial ~\gamma _{n}\left( \underline{\theta }\right) }{\partial
~\theta _{m}}=\gamma _{n}\left( \underline{\theta }\right) ~\left\{ \delta
_{nm}~\sum_{\ell =1,~\ell \neq n}^{N}\left[ \cot \left( \theta _{n}-\theta
_{\ell }\right) \right] -\left( 1-\delta _{nm}\right) ~\cot \left( \theta
_{n}-\theta _{m}\right) \right\} ~,
\end{equation}%
and likewise%
\begin{equation}
\frac{\partial ~\rho _{n}\left( \underline{\theta }\right) }{\partial
~\theta _{m}}=\rho _{n}\left( \underline{\theta }\right) ~\left\{ \delta
_{nm}~\sum_{\ell =1,~\ell \neq n}^{N}\left[ \cot \left( \theta _{n}-\theta
_{\ell }\right) \right] -\left( 1-\delta _{nm}\right) ~\cot \left( \theta
_{n}-\theta _{m}\right) \right\} ~.
\end{equation}%
Thereby the $N$-body system gets characterized by the following, simpler set
of Newtonian equations of motion: 
\end{subequations}
\begin{equation}
\mu _{n}~\ddot{\theta}_{n}=\sum_{\ell =1,~\ell \neq n}^{N}\left[ \frac{\dot{%
\theta}_{n}~\left( \mu _{\ell }~\dot{\theta}_{\ell }+\eta _{\ell }\right)
+\left( \mu _{n}~\dot{\theta}_{n}+\eta _{n}\right) ~\dot{\theta}_{\ell
}~\cos \left( \theta _{n}-\theta _{\ell }\right) }{\sin \left( \theta
_{n}-\theta _{\ell }\right) }\right] ~.  \label{WithTwoBodyForces}
\end{equation}

The differences among these two $N$-body systems, (\ref{WithManyBodyForces})
and (\ref{WithTwoBodyForces}), deserve to be emphasized: the $N$-body model (%
\ref{WithManyBodyForces}) involves \textit{many-body} forces, due to the
presence of the functions $\sigma _{n}\left( \underline{\theta }\right) $
and $\sigma _{\ell }\left( \underline{\theta }\right) $ in its right-hand
("forces") side; while the $N$-body model (\ref{WithTwoBodyForces}) only
involves \textit{two-body} forces. Both systems can be integrated once,
corresponding to the transition from their $N$ \textit{second-order}
Newtonian equations of motion to the corresponding $N$ \textit{first-order}
ODEs (\ref{hthetadot}). On the other hand, as we show below, only the first
of these two \textit{integrable} systems is \textit{solvable}.

Indeed, for the first system (but not for the second!), the $N$ first-order
ODEs (\ref{hthetadot}) are \textit{uncoupled}, reading simply, via (\ref%
{rhogammaconstants}), 
\begin{subequations}
\label{ODEsForThetan}
\begin{equation}
\mu _{n}~\dot{\theta}_{n}=-\eta _{n}+\sum_{m=1}^{N}\left[ h_{m}~s_{m}\left(
\theta _{n}\right) \right] ~,
\end{equation}%
or, equivalently (see (\ref{seeds}))%
\begin{equation}
\mu _{n}~\exp \left[ \left( N+1\right) ~i~\theta _{n}\right] ~\dot{\theta}%
_{n}=-\eta _{n}~\exp \left[ \left( N+1\right) ~i~\theta _{n}\right]
+\sum_{m=1}^{N}\left[ h_{m}~\exp \left( 2~m~i~\theta _{n}\right) \right] ~,
\end{equation}%
where the $N$ quantities $h_{n}$ are explicitly known in terms of the $2N$
initial data $\theta _{n}\left( 0\right) $, $\dot{\theta}_{n}\left( 0\right) 
$ (via (\ref{hn}), (\ref{rhogammaconstants}) and (\ref{qn}): see Appendix B).

These first-order ODEs can be integrated; we confine the relevant
developments to Appendix B.

Although the technique to manufacture these two \textit{solvable} and 
\textit{integrable} $N$-body problems, (\ref{WithManyBodyForces}) and (\ref%
{WithTwoBodyForces}), is \textit{not} new \cite{C2001}, these models are, to
the best of our knowledge, themselves \textit{new}; hence a detailed
discussion of the actual behavior of these systems has not yet been done. In
the present paper we limit our consideration to pointing out how these
models can be reformulated to describe the evolution of $N$ points whose
positions on a plane are characterized by $N$ \textit{unit} 2-vectors $\vec{r%
}_{n}\left( t\right) $, see the notation introduced in Subsection 2.1. To
this end one utilizes the formulas (\ref{rndotdot}), (\ref{rndotrm}), (\ref%
{rmwedgerndotzhat}) and the relevant ones among those conveniently collected
in Appendix A. And it is plain that one thereby obtains the two models (\ref%
{ManyBodyForcesModel}) and (\ref{TwoBodyForcesOnCircle}).

\subsection{Solvable models on the circle manufactured by reinterpreting
known solvable models}

In this section we tersely indicate how to obtain the two models (\ref{2a})
and (\ref{2b}).

The \textit{first model} obtains from the $N$-body system characterized by
the following Newtonian equations of motion (with velocity-independent
two-body forces): 
\end{subequations}
\begin{equation}
\ddot{\theta}_{n}=g^{2}~\sum_{\ell =1,~\ell \neq n}^{N}\left[ \frac{\cos
\left( \theta _{n}-\theta _{\ell }\right) }{\sin ^{3}\left( \theta
_{n}-\theta _{\ell }\right) }\right] ~.  \label{Suth}
\end{equation}%
Here $g$ is an arbitrary "coupling constant", and the rest of the notation
is, we trust, clear.

This is a well-known \textit{solvable} many-body problem, generally
associated with the name of Bill Sutherland, who was the first to show the
possibility to treat this $N$-body problem by exact methods (originally in a
quantal context \cite{S}); its treatment in a classical (Hamiltonian)
context is provided in several textbooks, see for instance \cite{P1990} \cite%
{C2001} \cite{S2004}.

It is plain that the model (\ref{2a}) is merely the transcription of this
model via the notation of Subsection 2.1.

The \textit{second model} obtains from the $N$-body system characterized by
the following Newtonian equations of motion (with velocity-dependent
one-body and two-body forces):%
\begin{equation}
\ddot{\theta}_{n}=g_{0}+g_{1}~\dot{\theta}_{n}+\sum_{\ell =1,~\ell \neq
n}^{N}\left\{ \left[ 2~\dot{\theta}_{n}~\dot{\theta}_{\ell }+g_{2}~\left( 
\dot{\theta}_{n}+\dot{\theta}_{\ell }\right) +g_{3}\right] ~\cot \left(
\theta _{n}-\theta _{\ell }\right) \right\} ~.
\end{equation}%
Here $g_{0},$ $g_{1},$ $g_{2}$ and $g_{3}$ are $4$ arbitrary coupling
constants, and we again trust the rest of the notation to be clear.

This is also a well known \textit{solvable} model, see for instance eq.
(2.3.5-12) on page 199 of \cite{C2001}.

And it is again plain that the model (\ref{2b}) is merely the transcription
of this model via the notation of Subsection 2.1 and Appendix A.

\subsection{How to manufacture $N$-body problems with \textit{angles} as
dependent variables}

In the preceding subsection we have shown how certain $N$-body models with
dependent variables naturally interpretable as \textit{angles} can be
reformulated as $N$-body models describing the time evolution on a plane of
particles \textit{constrained to move on a circle}. In this subsection we
indicate how, via a simple change of dependent variables, essentially 
\textit{any} $N$-body model can be reformulated so that its dependent
variables can be interpreted as \textit{angles}, hence subsequently it can
also be reformulated (in fact in many ways) so that it describes the time
evolution of particles \textit{constrained to move on a plane circle}.

The trick to achieve this goal is quite elementary and general; we
illustrate it below via two examples.

Consider an $N$-body model in which the positions of the $N$
point-particles---moving in one-dimensional space---are identified by $N$
coordinates $z_{n}\equiv z_{n}\left( t\right) ,$ and perform the change of
dependent variables by positing, say,%
\begin{equation}
z_{n}\left( t\right) =\tan \left[ \theta _{n}\left( t\right) \right] ~.
\label{ZnTanThetan}
\end{equation}

\textit{Remark 3.3.1}. Of course this assignment defines $\theta _{n}\left(
t\right) $ only $\func{mod}\left( \pi \right) $; and clearly many other
assignments could be instead made---different but having an analogous
effect, such as $z_{n}=1/\sin \left( 2\theta _{n}\right) $, or $z_{n}=\tan
^{3}\theta _{n}$, etc. . $\blacksquare $

In the \textit{first example} we take as point of departure the $N$-body
problem characterized by the Newtonian equations of motion 
\begin{subequations}
\begin{equation}
\ddot{z}_{n}=-4~z_{n}+g^{2}~\sum_{\ell =1,~\ell \neq n}^{N}\left[ \left(
z_{n}-z_{\ell }\right) ^{-3}\right] ~.  \label{Cal}
\end{equation}%
Here $g$ is an arbitrary (real) coupling constant. This is a well-known 
\textit{solvable} model (see for instance \cite{C2001}); it is \textit{%
isochronous}, all its solutions being \textit{completely periodic with
period }$\pi $, 
\begin{equation}
z_{n}\left( t\pm \pi \right) =z_{n}\left( t\right) ~.  \label{iso}
\end{equation}

Via the change of dependent variables (\ref{ZnTanThetan}) the equations of
motion (\ref{Cal}) become (as the diligent reader will easily verify,
utilizing if need be the identities reported in the last part of Appendix A) 
\end{subequations}
\begin{subequations}
\begin{eqnarray}
&&\ddot{\theta}_{n}=-2~\dot{\theta}_{n}^{2}~\tan \theta _{n}-4~\sin \theta
_{n}~\cos \theta _{n}  \notag \\
&&+g^{2}~\sum_{\ell =1,~\ell \neq n}^{N}\left[ \frac{\cos ^{5}\theta
_{n}~\sin ^{3}\theta _{\ell }}{\sin ^{3}\left( \theta _{n}-\theta _{\ell
}\right) }\right] ~.  \label{3a}
\end{eqnarray}

\textit{Remark 3.3.2}. This model of course hereditates the property of 
\textit{isochrony} of the model (\ref{Cal}) it has been obtained from:%
\begin{equation}
\theta _{n}\left( t\pm \pi \right) =\theta _{n}\left( t\right) ~~~\func{mod}%
\left( \pi \right) ~.~\blacksquare
\end{equation}

The next task is to transform these equations of motion, (\ref{3a}), into
equations of motion for points moving in the plane but constrained to stay
on a \textit{circle} of \textit{unit} radius centered at the origin. To
realize this goal one may now use the change of dependent variables from the 
\textit{angles} $\theta _{n}$ to the \textit{vectors} $\vec{r}_{n}$
described in Subsection 2.1, using if need be the identities reported in the
first part of Appendix A. And it is plain that in this manner one arrives at
the equations of motion (\ref{CalCircle}).

In the \textit{second example} we take as point of departure the well-known 
\textit{solvable }$N$-body problem characterized by the following Newtonian
equations of motion (see eq. (2.3.4.2-1) on page 188 of \cite{C2001}): 
\end{subequations}
\begin{equation}
\ddot{z}_{n}=-z_{n}+\sum_{\ell =1,~\ell \neq n}^{N}\left( \frac{2~\dot{z}%
_{n}~\dot{z}_{\ell }+1}{z_{n}-z_{\ell }}\right) ~.  \label{Gold}
\end{equation}%
\textit{All} solutions of this model are \textit{multiply periodic}, being
(generally nonlinear) superpositions of the $N$ functions $b_{m}\left(
t\right) =\cos \left( \sqrt{m}~t+\beta _{m}\right) ,$ $m=1,...,N$ (with the $%
N$ phases $\beta _{m}$ depending on the initial data); for special initial
data only functions $b_{m}\left( t\right) $ with $m$ a \textit{%
squared-integer} contribute, yielding solutions \textit{completely periodic}
with period $2\pi $. \cite{C2001}

Via the change of dependent variables (\ref{ZnTanThetan}) equations of
motion (\ref{Gold}) become (as the diligent reader will easily verify,
utilizing again, if need be, the identities reported in the last part of
Appendix A)%
\begin{eqnarray}
&&\ddot{\theta}_{n}=-2~\dot{\theta}_{n}^{2}~\tan \theta _{n}-\sin \theta
_{n}~\cos \theta _{n}  \notag \\
&&+\cos \theta _{n}~\sum_{\ell =1,~\ell \neq n}^{N}\left[ \frac{2~\dot{\theta%
}_{n}~\dot{\theta}_{\ell }+\cos ^{2}\theta _{n}~\cos ^{2}\theta _{\ell }}{%
\cos \theta _{n}~\sin \left( \theta _{n}-\theta _{\ell }\right) }\right] ~.
\end{eqnarray}%
Then we transform these equations of motion into equations of motion for
points moving in the plane but constrained to stay on a \textit{circle} of 
\textit{unit} radius centered at the origin, by using again the change of
dependent variables from the \textit{angles} $\theta _{n}$ to the \textit{%
vectors} $\vec{r}_{n}$ described in Subsection 2.1 via---if need be---the
identities reported in the first part of Appendix A. And it is plain that in
this manner one arrives at the equations of motion (\ref{GoldCircle}).

\section{Outlook}

Our original motivation to undertake this line of research was the intention
to manufacture $N$-body problems amenable to exact treatments describing
motions on a sphere, or more generally on manifolds. We consider the results
reported in this paper as a modest first step in that direction. We also
believe that the actual behavior of the \textit{new} models reported in this
paper---see (\ref{ManyBodyForcesModel}) and (\ref{TwoBodyForcesOnCircle}%
)---shall eventually deserve a more detailed scrutiny than that provided in
Subsection 3.1.

\section{Appendix A: identities}

It is plain that the notation introduced in Subsection 2.1 entails the
following additional identities: 
\begin{subequations}
\begin{equation}
\overset{\cdot }{\vec{r}}_{n}\cdot \vec{r}_{n}=0~,~~~\overset{\cdot }{\vec{r}%
}_{n}\cdot \overset{\cdot }{\vec{r}}_{n}=\dot{\theta}_{n}^{2},~~~\left( \vec{%
r}_{n}\wedge \overset{\cdot }{\vec{r}}_{n}\right) \cdot \hat{z}=\dot{\theta}%
_{n}~,  \label{Ardotscalar}
\end{equation}

\begin{equation}
\overset{\cdot \cdot }{\vec{r}}_{n}\cdot \vec{r}_{n}=-\dot{\theta}%
_{n}^{2}~,~~~\overset{\cdot \cdot }{\vec{r}}_{n}\cdot \left( \hat{z}\wedge 
\vec{r}_{n}\right) =\ddot{\theta}_{n}~,  \label{Ardotdotscalar}
\end{equation}%
\begin{equation}
\overset{\cdot }{\vec{r}}_{n}\cdot \vec{r}_{m}=-\dot{\theta}_{n}~\sin \left(
\theta _{n}-\theta _{m}\right) ~,  \label{Arndotrmscal}
\end{equation}%
\begin{equation}
\overset{\cdot }{\vec{r}}_{n}\cdot \overset{\cdot }{\vec{r}}_{m}=\dot{\theta}%
_{n}~\dot{\theta}_{m}~\cos \left( \theta _{n}-\theta _{m}\right) ~;
\label{Arndotrmdotscal}
\end{equation}%
\end{subequations}
\begin{subequations}
\begin{equation}
\hat{z}\wedge \overset{\cdot }{\vec{r}}_{n}=-\dot{\theta}_{n}~\vec{r}_{n}~,
\label{Azhatvectrdot}
\end{equation}

\begin{equation}
\hat{z}\wedge \overset{\cdot \cdot }{\vec{r}}_{n}=-\ddot{\theta}_{n}~\vec{r}%
_{n}-\dot{\theta}_{n}^{2}~\hat{z}\wedge \vec{r}_{n}~;
\label{Azhatvectrdotdot}
\end{equation}%
\end{subequations}
\begin{subequations}
\begin{equation}
\left( \overset{\cdot }{\vec{r}}_{n}\wedge \vec{r}_{m}\right) \cdot \hat{z}=-%
\dot{\theta}_{n}~\cos \left( \theta _{n}-\theta _{m}\right) ~,
\label{Arndotrmvect}
\end{equation}%
\begin{equation}
\left( \overset{\cdot }{\vec{r}}_{n}\wedge \overset{\cdot }{\vec{r}}%
_{m}\right) \cdot \hat{z}=-\dot{\theta}_{n}~\dot{\theta}_{m}~\sin \left(
\theta _{n}-\theta _{m}\right) ~.  \label{Arndotrmdotvect}
\end{equation}

We also display here some relations among the time-dependent "coordinates" 
\end{subequations}
\begin{subequations}
\begin{equation}
z_{n}\equiv z_{n}\left( t\right) =\tan \theta _{n}\left( t\right) ~,
\label{Axn}
\end{equation}%
and the "angles" $\theta _{n}\equiv \theta _{n}\left( t\right) $:%
\begin{equation}
z_{n}-z_{m}=\frac{\sin \left( \theta _{n}-\theta _{m}\right) }{\cos \theta
_{n}~\cos \theta _{m}}~,~~~\frac{1}{z_{n}-z_{m}}=\frac{\cos \theta _{n}~\cos
\theta _{m}}{\sin \left( \theta _{n}-\theta _{m}\right) }~;
\label{Aznminuszm}
\end{equation}%
\end{subequations}
\begin{equation}
\dot{z}_{n}=\frac{\dot{\theta}_{n}}{\cos ^{2}\theta _{n}}~,~~~\dot{z}%
_{n}~z_{m}=\frac{\dot{\theta}_{n}~\sin \theta _{m}}{\cos ^{2}\theta
_{n}~\cos \theta _{m}}~,~~~\dot{z}_{n}~\dot{z}_{m}=\frac{\dot{\theta}_{n}~%
\dot{\theta}_{m}}{\cos ^{2}\theta _{n}~\cos ^{2}\theta _{m}}~;
\label{Azndot}
\end{equation}%
\begin{subequations}
\label{Azndotzm}
\begin{equation}
\frac{\dot{z}_{n}+\dot{z}_{m}}{z_{n}-z_{m}}=\frac{\dot{\theta}_{n}~\cos
^{2}\theta _{m}+\dot{\theta}_{m}~\cos ^{2}\theta _{n}}{\cos \theta _{n}~\cos
\theta _{m}~\sin \left( \theta _{n}-\theta _{m}\right) }~,
\end{equation}%
\begin{equation}
\frac{\dot{z}_{n}~z_{m}+\dot{z}_{m}~z_{n}}{z_{n}-z_{m}}=\frac{\dot{\theta}%
_{n}~\sin \theta _{m}~\cos \theta _{m}+\dot{\theta}_{m}~\sin \theta
_{n}~\cos \theta _{n}}{\cos \theta _{n}~\cos \theta _{m}~\sin \left( \theta
_{n}-\theta _{m}\right) }~,
\end{equation}%
\begin{equation}
\frac{\dot{z}_{n}~\dot{z}_{m}}{z_{n}-z_{m}}=\frac{\dot{\theta}_{n}~\dot{%
\theta}_{m}}{\cos \theta _{n}~\cos \theta _{m}~\sin \left( \theta
_{n}-\theta _{m}\right) }~;
\end{equation}%
\end{subequations}
\begin{equation}
\ddot{z}_{n}=\frac{\ddot{\theta}_{n}}{\cos ^{2}\theta _{n}}+\frac{2~\dot{%
\theta}_{n}^{2}~\sin \theta _{n}~}{\cos ^{3}\theta _{n}}=\frac{\ddot{\theta}%
_{n}+2~\dot{\theta}_{n}^{2}~\tan \theta _{n}}{\cos ^{2}\theta _{n}}~.
\label{Azndotdot}
\end{equation}

\section{Appendix B: solution of the system (\protect\ref{ODEsForThetan})}

In this Appendix we indicate how the initial-value problem of the system of $%
N$ (decoupled)\ first-order ODEs (\ref{ODEsForThetan}) is solved.

Let us, for notational convenience, make here the following change of
variables: 
\begin{subequations}
\begin{equation}
\zeta _{n}\left( t\right) =\exp \left[ i~\theta _{n}\left( t\right) \right]
~,  \label{Bzn}
\end{equation}%
entailing%
\begin{equation}
\dot{\zeta}_{n}\left( t\right) =i~\dot{\theta}_{n}\left( t\right) ~\exp %
\left[ i~\theta _{n}\left( t\right) \right] ~.
\end{equation}

We then use the relation (\ref{Bzn}) to rewrite the equations of motion (\ref%
{ODEsForThetan}) as follows: 
\end{subequations}
\begin{equation}
\mu ~\zeta ^{N}~\dot{\zeta}=i~\left[ -\eta ~\zeta
^{N+1}+\sum_{m=1}^{N}\left( h_{m}~\zeta ^{2m}\right) \right] ~.  \label{ODE}
\end{equation}

\textit{Remark B.1}. Let us emphasize that, in the last formula and below
(in this Appendix B), as a notational simplification, we \textit{omit} to
indicate explicitly the time-dependence of the dependent variable $\zeta
_{n}\equiv \zeta _{n}\left( t\right) $, as well as its dependence on the
index $n$; and likewise the dependence on this index $n$ of the parameters $%
\mu _{n}$ and $\eta _{n}$. $\blacksquare $

The ODE (\ref{ODE}) can clearly be solved by the following quadrature:%
\begin{equation}
\dint\limits_{\zeta \left( 0\right) }^{\zeta \left( t\right) }d\xi ~\xi
^{N-2}~\left\{ -\eta ~\xi ^{N-1}+\sum_{m=1}^{N}\left[ h_{m}~\xi ^{2\left(
m-1\right) }\right] \right\} ^{-1}=\frac{i~t}{\mu }~.  \label{zt}
\end{equation}

To perform the integration it is convenient to introduce the $2\left(
N-1\right) $ zeros $\xi _{j}$ of the polynomial of degree $2\left(
N-1\right) $ appearing in the denominator of the integrand, 
\begin{subequations}
\begin{equation}
-\eta ~\xi ^{N-1}+\sum_{m=1}^{N}\left[ h_{m}~\xi ^{2\left( m-1\right) }%
\right] =h_{N}~\dprod\limits_{j=1}^{2\left( N-1\right) }\left( \xi -\xi
_{j}\right) ~,
\end{equation}%
and then the $2\left( N-1\right) $ "residues" $\phi _{j}$ defined by setting%
\begin{equation}
\left\{ -\eta ~\xi ^{N-1}+\sum_{m=1}^{N}\left[ h_{m}~\xi ^{2\left(
m-1\right) }\right] \right\} ^{-1}=h_{N}^{-1}~\dsum\limits_{j=1}^{2\left(
N-1\right) }\left( \frac{\phi _{j}}{\xi -\xi _{j}}\right) ~.  \label{Res}
\end{equation}%
Note that these formulas imply that the computation of, firstly, the $%
2\left( N-1\right) $ zeros $\xi _{j},$ and, secondly, the $2\left(
N-1\right) $ residues $\phi _{j},$ is a purely \textit{algebraic} task
(although not one that can be analytically performed for $N\geq 3$); hence
these quantities can in principle be considered known functions of the
parameter $\eta $ (from which they inherit a dependence on the index $n$,
see \textit{Remark B.1}) and of the $N$ constants of motion $h_{m}$. As for
these $N$ quantities $h_{m}$ (which are of course independent of the index $n
$) they are---in the context of the \textit{initial-value} problem for the
dynamical system (\ref{WithManyBodyForces})---explicitly given by the
formulas (\ref{hn}) at $t=0$ (let us reiterate that these expressions of the 
$N$ constants of motion $h_{m}$ are valid throughout the time evolution, and
of course, in particular, at the \textit{initial} time $t=0$).

The final step is to perform the integration in the left-hand side of (\ref%
{zt}). Via (\ref{Res}) the key ingredient to do so is the formula 
\end{subequations}
\begin{eqnarray}
&&\dint\limits_{\zeta _{0}}^{\zeta }d\xi ~\frac{\xi ^{N-2}}{\xi -\xi _{0}}%
=\dint\limits_{\zeta _{0}-\xi _{0}}^{\zeta -\xi _{0}}d\xi ~\frac{\left( \xi
+\xi _{0}\right) ^{N-2}}{\xi }  \notag \\
&=&\dint\limits_{\zeta _{0}-\xi _{0}}^{\zeta -\xi _{0}}d\xi ~\sum_{k=0}^{N-2}
\left[ \binom{N-2}{k}~\xi ^{k-1}~\xi _{0}^{N-2-k}\right]   \notag \\
&=&\xi _{0}^{N-2}~\log \left( \frac{\zeta -\xi _{0}}{\zeta _{0}-\xi _{0}}%
\right) +\sum_{k=1}^{N-2}\left\{ \binom{N-2}{k}~\frac{\xi _{0}^{N-2-k}}{k}~%
\left[ \left( \zeta -\xi _{0}\right) ^{k}-\left( \zeta _{0}-\xi _{0}\right)
^{k}\right] \right\} ~.  \notag \\
&&
\end{eqnarray}

\end{document}